\documentclass[10pt, journal]{IEEEtran}
\IEEEoverridecommandlockouts
\usepackage{cite}
\usepackage{amsmath,amssymb,amsfonts}
\usepackage{algorithmic}
\usepackage{graphicx}
\usepackage{textcomp}
\usepackage{xcolor}
\usepackage{breqn}
\usepackage{algorithm,algorithmic}
\usepackage{algorithm,algorithmic,amsbsy,amsmath,amssymb,epsfig,bbm,mathrsfs,fancyhdr,fancyvrb,url}
\usepackage{amssymb}
 
\usepackage{amsthm}
\usepackage{adjustbox,lipsum}
\usepackage{amsfonts}
\usepackage{epsfig}
\usepackage{amssymb}
\usepackage{amsmath}
\usepackage{cite}
\usepackage{subfigure}
\usepackage{multirow}
\usepackage{rotating}
\usepackage{graphicx}
\usepackage{tabularx}
\usepackage{array}
\usepackage{anyfontsize}
\usepackage{color,soul}
\usepackage{graphicx}
\usepackage{blindtext}
\usepackage{authblk}

\usepackage{amsmath}

\usepackage{stfloats}
\usepackage{verbatim}
\usepackage{balance}

\def\BibTeX{{\rm B\kern-.05em{\sc i\kern-.025em b}\kern-.08em
    T\kern-.1667em\lower.7ex\hbox{E}\kern-.125emX}}
\begin{document}

\title{RIS Meets Aerodynamic HAPS:\\
A Multi-objective Optimization Approach}
\author{Arman Azizi\thanks{The authors are with the Department of Electronic and Electrical Engineering, Trinity College Dublin, Dublin, D02 PN40 Ireland (e-mail: azizia@tcd.ie;  arman.farhang@tcd.ie). This publication has emanated from research supported in part by grants from Science Foundation Ireland under Grant numbers 18/CRT/6222, 13/RC/2077\textunderscore P2, 19/FFP/7005(T) and 21/US/3757. For the purpose of Open Access, the author
has applied a CC BY public copyright licence to any Author Accepted Manuscript version arising
from this submission.}, \textit{Student Member, IEEE}, Arman Farhang, \textit{Senior Member, IEEE} \vspace{-9mm}}
\maketitle
\begin{abstract}
In this letter, we propose a novel network architecture for integrating terrestrial and non-terrestrial networks (NTNs) to establish connection between terrestrial ground stations which are unconnected due to blockage. We propose a new network framework where reconfigurable intelligent surface (RIS) is mounted on an aerodynamic high altitude platform station (HAPS), referred to as aerodynamic HAPS-RIS. This can be one of the promising candidates among non-terrestrial RIS (NT-RIS) platforms. We formulate a mathematical model of the cascade channel gain and time-varying effects based on the predictable mobility of the aerodynamic HAPS-RIS. We propose a multi-objective optimization problem for designing the RIS phase shifts to maximize the cascade channel gain while forcing the Doppler spread to zero, and minimizing the delay spread upper bound.
Considering an RIS reference element, we find a closed-form solution to this optimization problem based on the Pareto optimality of the aforementioned objective functions. Finally, we evaluate and show the effective performance of our proposed closed-form solution through numerical simulations.
\end{abstract}
\vspace{-1mm}
\vspace{-4mm}
\section{Introduction}
One of the most important targets in sixth generation wireless networks (6G) is the provision of ubiquitous connectivity. This aim can be attained by integration of terrestrial and non-terrestrial networks (NTNs), \cite{3GPP release}. To this end, reconfigurable intelligent surface (RIS) can be exploited to boost the channel gain by creating a multi-path environment. Non-terrestrial RIS (NT-RIS) is an intelligent intermediate reflection layer, where RIS is mounted on a non-terrestrial platform to connect the unconnected terrestrial infrastructures. 
\textcolor{black}{Extensive research has been conducted to address the benefits of adopting NT-RIS in wireless networks, see \cite{NTN-RIS-alouini, NTN-RIS-jamalipour, Halim-HAPS-UAV-SAT, Halim-link-budget} and the references therein.} 
In practical cases, high altitude platform station (HAPS)-RIS is one of the promising candidates to be exploited for NT-RIS compared to other non-terrestrial platforms such as satellite-RIS and unmanned aerial vehicle (UAV)-RIS, \cite{Halim-HAPS-UAV-SAT,Halim-link-budget}. 
\textcolor{black}{HAPS operates at much higher altitude which leads to establishing line of sight (LoS) dominated connection and a much wider coverage area compared to UAV. Furthermore, HAPS is much larger than UAV so that RIS with a large number of elements can be mounted on it \cite{Halim-HAPS-UAV-SAT,NTN-RIS-alouini}.}
\textcolor{black}{The advantages of exploiting RIS over relay is well articulated in \cite{NTN-RIS-alouini}, e.g., lower-cost, simpler hardware, shorter transmission delay, less power consumption, and longer communication duration. In \cite{RIS-Relay}, the authors prove that if the RIS is large enough it can beat the relay in terms of energy-efficiency. Due to the large size of HAPS, large number of RIS elements can be mounted on HAPS, and hence, RIS is the better option.
Even if a large number of RIS elements are deployed, the HAPS's payload is light due to the thin and lightweight materials from which the RIS elements are manufactured \cite{NTN-RIS-alouini}.} \textcolor{black}{From the perspective of HAPS mobility, there are two types of HAPSs, aerostatic and aerodynamic, \cite{Halim-HAPS-Types}. 
The investigation of HAPS-RIS communications is still in its infancy. The existing literature on this topic is mostly focused on aerostatic HAPS-RIS, \cite{Halim-HAPS-UAV-SAT, Halim-link-budget, HAPS-RIS-ICC, HAPS-RIS-efficient}, while the aerodynamic HAPS-RIS is left as an open research topic. \textcolor{black}{The necessity of research on this direction has been emphasised in \cite{High Mobility RIS}.}
The advantages of exploiting aerodynamic over aerostatic HAPS in wireless networks are well articulated in \cite{Halim-HAPS-Types}, e.g., low-cost and swift deployment, and high resilience to turbulence.
These features make aerodynamic HAPS a promising candidate technology in the move towards integration of terrestrial and non-terrestrial networks, \cite{Halim-HAPS-Types}.}
However, high mobility of aerodynamic HAPS leads to time-varying channel effects. Accordingly, the main research question that arises is 
``\textit{{Can aerodynamic HAPS-RIS bring connectivity to the unconnected ground stations in presence of time-varying channel?}}". 
 
 There exist a number of works in the literature that consider
 RIS-based networks in the presence of time-varying channel, which can be classified into two groups where RIS is fixed, \cite{CE-RIS-TV, RIS-Roadside-Rzhang, Emil WCL}, or mobile, \cite{CE-MobileRIS-TV, CE-MobileRIS-TWC, RIS-SAT-JSAC}.
\textcolor{black}{Our proposed network architecture in this letter falls under the area of the latter one, where the RIS is mobile.
In \cite{CE-MobileRIS-TV} and \cite{CE-MobileRIS-TWC}, the authors present efficient Doppler shift mitigation methods, including transmission protocol and RIS phase shift control, where both of RIS and user equipment are deployed in a high-mobility terrestrial vehicle. The main difference between \cite{CE-MobileRIS-TV} and \cite{CE-MobileRIS-TWC}, is the design of the transmission protocol.
In \cite{RIS-SAT-JSAC}, the authors present a cooperative passive beamforming and distributed channel estimation to maximize the overall channel gain between an RIS-aided low-earth orbit satellite and a ground node. While the main focus of \cite{CE-MobileRIS-TV, CE-MobileRIS-TWC, RIS-SAT-JSAC} is channel estimation, to the best of our knowledge, there is no existing work which geometrically formulates all the channel metrics and time-varying effects based on predictive mobility of RIS, which can play a vital role in reducing the computational complexity. 
Furthermore, the authors in \cite{CE-MobileRIS-TV, CE-MobileRIS-TWC, RIS-SAT-JSAC} only consider one side of the cascade channel to be time-varying, while in this letter we investigate the case where both sides of the cascade channel are time-varying.} 
\\
To summarize, this letter addresses the aforementioned gaps in the literature with the ensuing contributions: (i) We introduce a \textit{novel network architecture} for NT-RIS assisted networks. We propose a new system model where RIS is mounted on aerodynamic HAPS to connect the unconnected terrestrial ground stations in emergency situations thanks to significant features of aerodynamic HAPS. 
 (ii) We \textit{mathematically model the mobility pattern of each RIS element} based on the dimensions of the RIS and the RIS elements, and the predictive trajectory of the aerodynamic HAPS-RIS. Next, we obtain a geometrical model for all the channel metrics and time-varying effects. To the best of our knowledge, there is no work which geometrically models the the mobility profile of a mobile RIS based on these parameters. (iii) We propose a multi-objective optimization problem in which the objective functions are the channel gain, the delay spread upper bound and the Doppler spread. We \textit{obtain a closed-from solution} for the RIS phase shifts, by introducing a reference RIS element, adopting Pareto optimality. \textcolor{black}{The obtained closed-form is based on the known locations of Tx and Rx, and the known time-varying position of the aerodynamic HAPS-RIS, thanks to knowing its mobility pattern. This leads to practical and simple implementation as the RIS phase shifts can be efficiently calculated by the onboard processing unit on HAPS.}
\vspace{-3mm}
\section{System Model and Problem Formulation}
\begin{figure}
\centering
\includegraphics[scale=0.24]{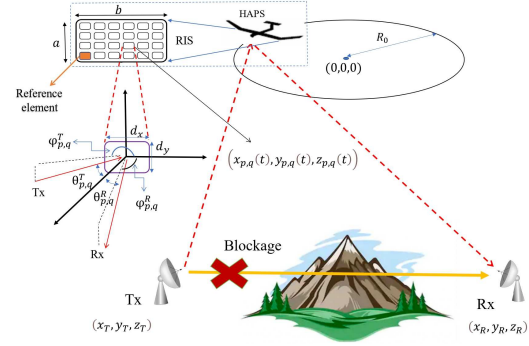}
        \vspace{-3mm}
    \caption{Proposed network architecture based on aerodynamic HAPS-RIS.}
    \vspace{-2mm}
    \label{sys}
\end{figure}

\begin{figure}[t!]
\centering
\includegraphics[scale=0.265]{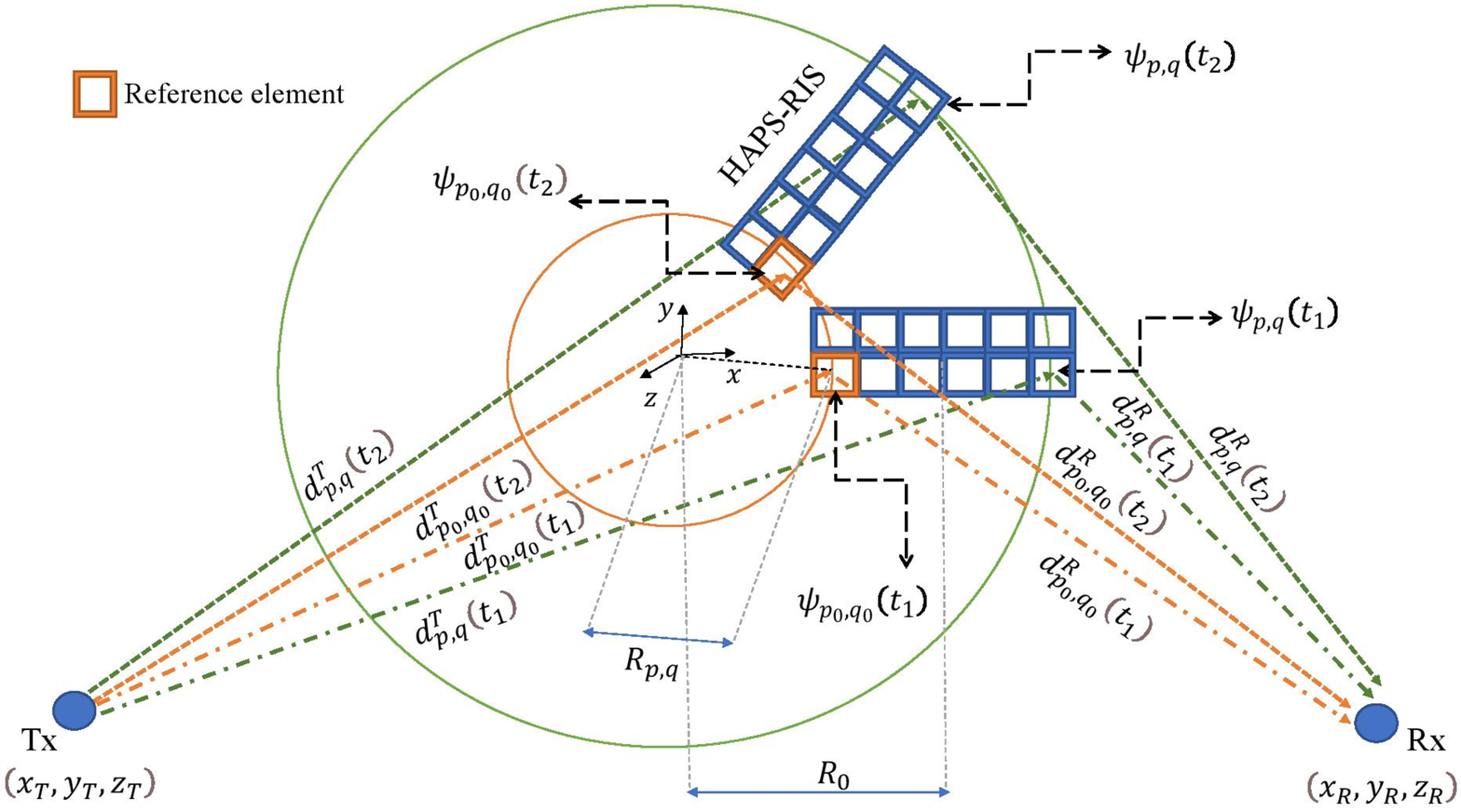}
    \caption{\textcolor{black}{Geometrical mobility pattern of RIS elements based on the predictable mobility of the aerodynamic HAPS.}}
    \label{RIS mobility pattern}
\end{figure}

\textcolor{black}{In this letter, we consider a heavy blockage scenario where the link between the terrestrial transmitter (Tx) and receiver (Rx) is blocked. 
We consider the network architecture in Fig.~\ref{sys}, exploiting aerodynamic HAPS-RIS to connect the unconnected ground stations.} We consider the RIS to be a rectangle with the length $a$ and the width $b$, which is located on the bottom of the HAPS in the $xy$-plane.
\textcolor{black}{$d_{x}$ and $d_{y}$ are the dimensions of each RIS element, which are in the range of $[\frac{\lambda_{\rm{c}}}{10}, \frac{\lambda_{\rm{c}}}{5}]$ where $\lambda_{\rm{c}}$=$\frac{c_{0}}{f_{\rm{c}}}$ is the carrier wavelength \cite{Emil-RIS-Dimensions}. $f_{\rm{c}}$ is the carrier frequency and $c_{0}$ is the speed of light.} 
The RIS consists of $P=\left\lceil \frac{a}{d_{x}} \right\rceil$ columns and $Q=\left\lceil \frac{b}{d_{y}} \right\rceil$ rows of reflecting elements. 
\textcolor{black}{The aerodynamic HAPS has a circular movement in the stratosphere} with radius $R_0$ centered at the origin of the Cartesian coordinate system and the velocity $v$. 
\textcolor{black}{ It is not practical to consider different trajectories for the aerodynamic HAPS, like what is expected for UAVs. It is vital to consider the circular trajectory for the aerodynamic HAPS, which leads to quasi-stationary position, that brings resilience to turbulence \cite{Halim-HAPS-Types}.}
\textcolor{black}{As the aerodynamic HAPS is moving with high speed, both sides of the cascade channel for each RIS element are time-varying. This can be clearly observed in Fig.~\ref{RIS mobility pattern} which shows the geometrical mobility pattern of RIS elements based on the predictive mobility of the aerodynamic HAPS.}  \\
\textbf{\textit{Definition 1.}}
\textit{The geometrical mobility pattern of the RIS elements can be attained as a function of the predictive mobility of the aerodynamic HAPS, and the dimensions \textcolor{black}{$a$, $b$, $d_{x}$, and $d_{y}$} as  
$(x_{p,q}\left( t\right),y_{p,q}(t),z_{p,q}(t)  )=( R_{p,q}\cos ( \frac{vt}{R_{p,q}}+\alpha_{p,q}),R_{p,q}\sin (\frac{vt}{R_{p,q}}+\alpha_{p,q}),0)$
where
\vspace{-1mm}
\begin{equation}
\begin{array}{l} \hspace{-2mm}
   R_{p,q}=\sqrt{(R_{0}-\frac{a}{2}+( p-\frac{1}{2} ) d_{x})^{2}+     
     (-\frac{b}{2}+( q-\frac{1}{2} )  d_{y})^{2}},
\end{array}
\end{equation}
\begin{equation}
    \alpha_{p,q}=\arctan\left(\frac{-\frac{b}{2}+( q-\frac{1}{2} )  d_{y}}{R_{0}-\frac{a}{2}+( p-\frac{1}{2} ) d_{x}}\right).
\end{equation}}


\textcolor{black}{We consider a LoS dominated scenario for the links between Tx/Rx and HAPS-RIS as the aerodynamic HAPS-RIS flies in a high altitude, i.e., 20 km \cite{Halim-HAPS-Types}. Furthermore, in HAPS-RIS scenarios, the ground stations are considered as high directional antenna gain transceivers which leads to establishing strong and dominant LoS link \cite{HAPS-RIS-ICC, HAPS-RIS-efficient}.}
The Tx sends a passband signal $s_{p}\left( t\right)  =\sqrt{2} \Re \left\{ s\left( t\right)  \exp \left( j2\pi f_{\rm{c}}t\right)  \right\}=\frac{s\left( t\right)  \exp \left( j2\pi f_{\rm{c}}t\right)+s^{*}\left( t\right)  \exp \left( -j2\pi f_{\rm{c}}t\right)  }{\sqrt{2}}$  where $s(t)$ is the complex baseband signal with bandwidth $B/2$ which is modulated to the carrier frequency $f_{\rm{c}}$ satisfying $B\ll2f_{\rm{c}}$, \cite{emil mag}.
Thus, the received baseband signal can be shown as 
$r\left(t\right)=\sum\limits_{p=1}^{P}\sum\limits_{q=1}^{Q} \Gamma_{p,q}(t) \exp({-j2\pi f_{\rm{c}}\tau_{p,q} (t) } -j\psi_{p,q}(t)) s( t-\tau_{p,q} (t)  -\frac{\psi_{p,q}(t)}{2\pi f_{\rm{c}}})  +n(t)$
where $\Gamma_{p,q}(t)$ is the cascade channel gain coefficient for the RIS element $(p,q)$ and $n(t)$ is the additive white Gaussian noise (AWGN). Additionally, $\psi_{p,q}(t)$ is the phase shift of the RIS element $(p,q)$. Using the Friis model, \cite{Friis}, $\Gamma_{p,q}(t)$ is the multiplication of the ground to air and air to ground amplitude gains as
\vspace{-1.5mm}
\begin{equation}\label{Friis}
    \Gamma_{p,q} \left( t\right)  =\frac{\lambda^{2}_{c} }{16\pi^{2} \prod\limits_{S } d^{S }_{p,q}\left( t\right)  } \sqrt{\prod\limits_{S } g^{p,q}_{S }\left( t\right)  \prod\limits_{S } g^{S }_{p,q}\left( t\right)  }, 
\end{equation}
where $S \in\{\rm{T,R}\}$ represents the Tx/Rx. The distance between the RIS element $(p,q)$ and the Tx/Rx can be calculated as $d^{S  }_{p,q}(t)=
     \sqrt{( x_{p,q}(t)  -x_{S  })^{2}  +(y_{p,q}( t)  -y_{S  })^{2}  +( z_{p,q}(t)-z_{S  })^{2}}$.~Moreover,
$g^{S }_{{}p,q}\left( t\right)$ is the antenna gain of RIS element $(p,q)$ to $S$, which can be a function of $\theta^{S }_{p,q} \left( t\right)  \in \left[ 0,\pi \right]  $ and $\varphi^{S }_{p,q} \left( t\right)  \in \left[ 0,2\pi \right]$. We consider that 
$g^{S }_{{}p,q}\left( t\right)$ is zero for $\theta^{S }_{p,q} \left( t\right)  \in \left[ \frac{\pi}{2},\pi \right]  $. The term $\theta^{S }_{p,q}(t)=\arccos ( \frac{z_{S  }-z_{p,q}(t)}{d^{S }_{p,q}( t) }) $ is the elevation angle from the RIS element $(p,q)$ to $S $. The term $\varphi^{S  }_{p,q} \left( t\right)  =\arctan ( \frac{y_{S  }-y_{p,q}\left( t\right)}{x_{S  }-x_{p,q}\left( t\right)} )$ is the azimuth angle from the RIS element $(p,q)$ to $S$. Furthermore, $g^{p,q}_{{}S }\left( t\right)$ is the antenna gain of the Tx/Rx to/from the RIS element $(p,q)$. The terms $\theta^{p,q}_{S } \left( t\right)$ and 
$\varphi^{p,q}_{S } \left( t\right)$ are the angle of elevation and azimuth from $S$ to the RIS element $(p,q)$, respectively.
$\tau_{p,q}(t)$ is the cascade path delay for the RIS element $(p,q)$, which can be formulated as $\tau_{p,q} (t)=\frac{\sum\limits_{S } d^{S }_{p,q}\left( t\right)  }{c_{0}} $.
\\
\textcolor{black}{ The instantaneous cascade channel gain as the ratio between the received power, $P_{\rm{R}}(t)$, and the time-invariant transmit power, $P_{\text{T}}$, can be obtained as }
\vspace{-3mm}
\begin{equation}\label{received power}
   \frac{P_{\rm{R}}(t)}{P_{\text{T}}}= \left|\sum\limits_{p=1}^{P}\sum\limits_{q=1}^{Q} \Gamma_{p,q}(t) \exp(-j2\pi f_{\rm{c}}\tau_{p,q}( t)
        -j\psi_{p,q}( t))\right|^{2},
\end{equation}
\textcolor{black}{The time-varying effects caused by the cascade paths through the RIS elements, i.e., the Doppler spread, $B_{\rm{Do}}(t)$, and the delay spread, $T_{\rm{De}}\left( t\right)$, can be obtained as \cite{Emil WCL, emil mag}}
 \vspace{-1mm}
\begin{equation}\label{doppler}\hspace{-2mm}
    \begin{array}{l}
         B_{\rm{Do}}(t)=
         f_{\rm{c}}\times\\
         \mathop {\max }\limits_{p,q,p^{\prime },q^{\prime }} \left| \frac{d}{dt} (\tau_{p,q} (t)  +  
          \frac{\psi_{p,q}\left( t\right)}{2\pi f_{\rm{c}}} ) -
          \frac{d}{dt}( \tau_{p^{\prime},q^{\prime}} \left( t\right) +\frac{\psi_{p^{\prime },q^{\prime }}\left( t\right)}{2\pi f_{\rm{c}}} )\right|,
    \end{array}
\end{equation}
\begin{equation}\label{eq: delay spread}
      T_{\rm{De}}\left( t\right) =\mathop {\max }\limits_{p,q} \{ \tau_{p,q} \left( t\right)  +\frac{\psi_{p,q} \left( t\right)  }{2\pi f_{\rm{c}}} \}  - 
      \mathop {\min }\limits_{p,q} \{ \tau_{p,q} \left( t\right)  +\frac{\psi_{p,q} \left( t\right)  }{2\pi f_{\rm{c}}} \}.
\end{equation}

\textcolor{black}{We consider a multi-objective optimization problem, including analogous objective functions as in \cite{Emil WCL}, to maximize \eqref{received power} while minimizing \eqref{doppler} and \eqref{eq: delay spread} simultaneously. Therefore, the main optimization problem can be formulated as }
\vspace{-2mm}
\begin{equation}\label{equation: main optimization problem}
\textcolor{black}{{\textbf{\rm{OP}}}_{\boldsymbol{}}: 
  \max_{\forall p,q,t:\  \textcolor{black}{\psi_{p,q} \left( t\right) \geq 0 }} \left[ \frac{P_{\rm{R} }\left( t\right)  }{P_{\text{T} }}, -B_{\rm{Do} }\left( t\right),   -T_{\rm{De} }\left( t\right)  \right].}
\end{equation}
\textcolor{black}{The feasible set ${\psi}_{p,q}(t) \geq 0$, originates from the causality requirement \cite{emil mag}.}
\textcolor{black}{We consider a mobile RIS where both sides of the cascade channel are time-varying while in \cite{Emil WCL} the RIS is fixed and only one side of the cascade channel is time-varying. Furthermore, we consider the link between Tx and Rx is blocked while in \cite{Emil WCL} the direct link is available. The adopted technique in \cite{Emil WCL} does not work for our proposed model to solve ${\textbf{\rm{OP}}}_{\boldsymbol{}}$. To tackle this issue, as can be seen in Fig. \ref{RIS mobility pattern}, we consider a  single RIS element as a reference with variable phase shift $\psi_{p_{0},q_{0}}(t)$, so that the other phase shifts can be obtained based on that. The cascade path through the reference element is called reference path.} 
\vspace{-3mm}
\section{Proposed RIS Phase Shift Design}\label{sec: RIS phase shift design}
 \textcolor{black} { 
 To find the optimal solution of ${\textbf{\rm{OP}}}_{\boldsymbol{}}$, let us consider the search space as the set $\boldsymbol{\Psi}$. Even if we relax the continuous RIS phase shifts to discrete ones with $M$ quantization levels, to simplify the problem, the search space has $M^{PQ}$ states. As this is a massive number for a large number of RIS elements, finding the optimal solution is intractable in terms of computational complexity. For large values of $M$, to get close to the continuous case, the search space $\boldsymbol{\Psi}$ becomes prohibitively large. Thus, it is evident that if the phase shifts are continuous like our proposed scenario, solving \eqref{equation: main optimization problem} is not affordable in terms of computational complexity. To tackle this issue, we find the Pareto optimal solution of ${\textbf{\rm{OP}}}_{\boldsymbol{}}$ in \textit{{Proposition 1}} by decomposing ${\textbf{\rm{OP}}}_{\boldsymbol{}}$ into ${\textbf{\rm{OP}}}_{\boldsymbol{1}}$ and ${\textbf{\rm{OP}}}_{\boldsymbol{2}}$.}\\
 \vspace{0mm}
 \begin{figure}
\centering
    \includegraphics[scale=0.26]{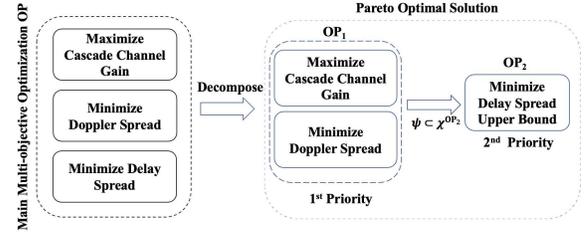}    
        \vspace{-3mm}
\caption{\textcolor{black}{The proposed Pareto Optimal Solution Method}}
    \vspace{-2mm}
    \label{pareto}
\end{figure}
\textcolor{black}{\textit{\textbf{Proposition 1.}} 
\textit{Let us decompose ${\textbf{\rm{OP}}}_{\boldsymbol{}}$ into ${\textbf{\rm{OP}}}_{\boldsymbol{1}}$ and ${\textbf{\rm{OP}}}_{\boldsymbol{2}}$ as
\begin{equation}\label{OP2}
\hspace{-1mm}
     {\textbf{\rm{OP}}}_{\boldsymbol{1}}: \forall {p,q\textcolor{black}{,t}}: \psi_{p,q}(t)  \in \arg \max_{\boldsymbol{\Psi}} \frac{P_{\rm{R}}\left( t\right)}{P_{\rm{T}}}  \cap
     \arg \min_{\boldsymbol{\Psi}} \ B_{\rm{Do}}(t),
\end{equation}
\begin{equation}\label{OP3}
\hspace{-2.2cm}
    {\textbf{\rm{OP}}}_{\boldsymbol{2}}: \forall p,q\textcolor{black}{,t}:\  \psi_{p,q} \left( t\right)  \in \arg \min_{\boldsymbol{\psi} \subset \boldsymbol{\chi}^{{\textbf{\rm{OP}}}_{{2}} } } T^{\rm{upp} }_{\rm{De} }\left( t\right).
\end{equation}
\textit{As ${\textbf{\rm{OP}}}_{\boldsymbol{1}}$ and ${\textbf{\rm{OP}}}_{\boldsymbol{2}}$ can not be optimized simultaneously, we consider that ${\textbf{\rm{OP}}}_{\boldsymbol{1}}$ has the higher priority order compared to ${\textbf{\rm{OP}}}_{\boldsymbol{2}}$, which is elaborated later in {\textit{Lemma 2}}.
\textcolor{black}{As can be seen in Fig.~\ref{pareto}, ${\textbf{\rm{OP}}}_{\boldsymbol{1}}$ optimizes $\frac{P_{\rm{R} }\left( t\right)  }{P_{\text{T} }}$ and $B_{\rm{Do}}(t)$ simultaneously. Let us consider all the possible solutions of ${\textbf{\rm{OP}}}_{\boldsymbol{1}}$ is the solution set $\boldsymbol{\chi}^{{\textbf{\rm{OP}}}_{{2}} }$ which is a feasible set for ${\textbf{\rm{OP}}}_{\boldsymbol{2}}$. In ${\textbf{\rm{OP}}}_{\boldsymbol{2}}$, we optimize the delay spread upper bound, $T^{\rm{upp} }_{\rm{De}}(t)$, over the feasible set $\boldsymbol{\psi} \subset \boldsymbol{\chi}^{{\textbf{\rm{OP}}}_{{2}}}$ resulting from solving ${\textbf{\rm{OP}}}_{\boldsymbol{1}}$.}
}}\textit{The Pareto optimal closed-form solution of} \eqref{equation: main optimization problem} \textit{is}
\begin{equation} \label{proposed solution}
    \psi_{p,q} \left( t\right)  =2\pi \text{mod} \left( f_{\rm{c}}\left( \tau_{p_{0},q_{0}} \left( t\right)  -\tau_{p,q} \left( t\right)  \right)  ,1\right),
\end{equation}
\textit{where $\text{mod}(\mu,\eta)$ is the remainder of the division of $\mu$ by $\eta$. \textcolor{black}{$\tau_{p_{0},q_{0}}(t)$ is the delay spread through the RIS reference element. \textcolor{black}{ There is no need to consider $\psi_{p,q} \left( t\right) \leq 2\pi$ as it is already satisfied in our closed-form solution.} }}
\begin{proof} The Pareto optimal solution can be attained based on lemma 1 and 2.
\end{proof}} 
\vspace{-2mm}
\textcolor{black}{\textit{\textbf{Lemma 1.}} 
\textit{\textcolor{black}{As the Doppler spread, $B_{\rm{Do}}(t)$ in \eqref{doppler}, is a function of RIS phase shifts, \textcolor{black}{using the following criterion in RIS phase shift design, this effect becomes zero.}}
\begin{equation}\label{doppler zero phase shift}\hspace{-0.3cm}
\begin{array}{l}
    \frac{d}{dt} \psi_{\textcolor{black}{\tilde{p},\tilde{q}}} \left( t\right)= \frac{d}{dt}\psi_{p_{0},q_{0}}(t)+
     2\pi f_{\rm{c}}\frac{d}{dt}\varpi_{\textcolor{black}{\tilde{p} ,\tilde{q}} }(t),
\end{array}
\end{equation}}
where $\varpi_{\textcolor{black}{\tilde{p} ,\tilde{q}} }(t)=\tau_{p_{0},q_{0}} \left( t\right)  -\tau_{\textcolor{black}{\tilde{p} ,\tilde{q}} } \left( t\right)$.}
\textcolor{black}{For brevity, RIS elements except the reference element are shown as $(\textcolor{black}{\tilde{p} ,\tilde{q}})$.}
\begin{proof}
The Doppler spread can be represented as 
\begin{equation}\label{DopplerSpread}\hspace{-2mm}
\begin{array}{l}
    B_{\rm{Do}}(t)=
    \max\{ B_{\rm{Do},1}(t),B_{\rm{Do},2}(t)\},
    \end{array}
\end{equation}
where the Doppler spread between the reference path and other cascade paths is
\begin{equation}\label{Bd1}\hspace{-7mm}
\begin{array}{l}
       B_{\rm{Do},1}(t)=f_{\rm{c}} \mathop {\max}\limits_{\textcolor{black}{\tilde{p} ,\tilde{q}}}| \frac{d}{dt} ( \tau_{\textcolor{black}{\tilde{p} ,\tilde{q}} }(t)  + \frac{\psi_{\textcolor{black}{\tilde{p} ,\tilde{q}} }(t)}{2\pi f_{\rm{c}}})
       -\\ ~~~~~~~~~~~~~~~~~~~~~~~~~~~~~\frac{d}{dt} ( \tau_{p_{0},q_{0}} ( t)+\frac{\psi_{p_{0},q_{0}}(t)}{2\pi f_{\rm{c}}})|,
\end{array}
\end{equation}
and the Doppler spread between the cascade paths except reference path is
\begin{equation}\label{Bd2}\hspace{-3mm}
    \begin{array}{l}
          B_{\rm{Do},2}(t)=
         f_{\rm{c}}\mathop {\max }\limits_{\textcolor{black}{\tilde{p} ,\tilde{q}, \tilde{p}^{\prime} ,\tilde{q}^{\prime}} } | \frac{d}{dt} (\tau_{\textcolor{black}{\tilde{p} ,\tilde{q}}} \left( t\right)  +  
         \frac{\psi_{\textcolor{black}{\tilde{p} ,\tilde{q}}}\left( t\right)}{2\pi f_{\rm{c}}}) 
          -\\ ~~~~~~~~~~~~~~~~~~~~~~~~~~~~~~~~~~\frac{d}{dt}(\tau_{\textcolor{black}{\tilde{p}^{\prime} ,\tilde{q}^{\prime}} } \left( t\right) +\frac{\psi_{\textcolor{black}{\tilde{p}^{\prime} ,\tilde{q}^{\prime}} }\left( t\right)}{2\pi f_{\rm{c}}})|.
    \end{array}
\end{equation}
In order to make the Doppler spread zero, we force both $B_{\rm{Do},1}$ and $B_{\rm{Do},2}$ to zero, which leads to \eqref{doppler zero phase shift}.
\end{proof}
\vspace{-1mm}
\textbf{\textit{Lemma 2.}}
\textit{ The Pareto optimal solution, \eqref{proposed solution}, optimizes \eqref{received power} and \eqref{doppler} simultaneously, and after that minimizes $T_{\rm{De}}^{\rm{upp}}(t)$.}
\vspace{-2mm}
\begin{proof}
After forcing Doppler spread to zero, we have a feasible set for ${\psi}_{p,q}(t)$ based on \eqref{doppler zero phase shift}.
First, we integrate \eqref{doppler zero phase shift} with respect to $t$ and substitute the result into \eqref{received power}. In the next step, in order to maximize the instantaneous cascade channel gain, all the terms of \eqref{received power} should have the same phase. Therefore, $\forall{p,q}$ choosing
\vspace{-2mm}
\textcolor{black}{\begin{equation}\label{max power phase shift}\hspace{-2mm}
 \begin{array}{l}\psi_{p,q} \left( t\right) = \\\begin{cases}\psi_{p_{0},q_{0}}(t)  &  p=p_{0},q=q_{0},\\ 2\pi f_{\rm{c}} \varpi_{\textcolor{black}{\tilde{p} ,\tilde{q}} }(t) + 2\pi \zeta_{\textcolor{black}{\tilde{p} ,\tilde{q}}} \left( t\right)+\psi_{p_{0},q_{0}}(t) & \text{Otherwise},\\  \end{cases} \end{array} 
\end{equation}}and $\zeta_{\textcolor{black}{\tilde{p} ,\tilde{q}}} \left( t\right)\in \mathbb{Z}$, maximize \eqref{received power}. It is clear that \eqref{received power} is the most important metric among the objective functions, which leads to maximizing signal-to-noise ratio. From \eqref{doppler zero phase shift} and \eqref{max power phase shift}, we see that \eqref{received power} and \eqref{doppler} can be simultaneously optimized, irrespective of the phase shift $\psi_{p_{0},q_{0}}(t)$. Due to the causality requirement, ${\psi}_{p,q}(t) \geq 0$, we can attain the upper bound of delay spread based on \eqref{eq: delay spread} as
\vspace{-2mm}
\begin{equation}\label{delay spread upper bound}
\begin{array}{l}
      T_{\rm{De}}^{\rm{upp}}(t) =\mathop {\max }\limits_{p,q} \{ \tau_{p,q}(t)  +\frac{\psi_{p,q} (t)  }{2\pi f_{\rm{c}}}\}  - 
      \mathop {\min }\limits_{p,q} \{ \tau_{p,q}( t)\} .
\end{array}
\end{equation}    
From \eqref{max power phase shift} and \eqref{delay spread upper bound}, it is obvious that there is no single solution for optimizing  ${\textbf{\rm{OP}}}_{\boldsymbol{1}}$ and ${\textbf{\rm{OP}}}_{\boldsymbol{2}}$ simultaneously. As \eqref{delay spread upper bound} is an increasing function in $\psi_{p,q}(t)$, zero phase shift is needed $\forall p,q,t$ to minimize \eqref{delay spread upper bound},  which is impossible according to \eqref{max power phase shift}. Instead, there are infinite non-inferior solutions, \cite{Emil-MOO}.
\textcolor{black}{By substituting \eqref{max power phase shift} into \eqref{delay spread upper bound}, the delay spread upper bound can be obtained based on the possible solutions of ${\textbf{\rm{OP}}}_{\boldsymbol{1}}$ as 
\begin{equation}\label{Delay Spread Upper Bound after OP2}
    \begin{array}{l}
       T^{\rm{upp}}_{\rm{De} }\left( t\right)  =\max \{ \tau_{p_{0 },q_{0}} \left( t\right)  +\frac{\psi_{p_{0 },q_{0 }} \left( t\right)  }{2\pi f_{\rm{c}}} ,\mathop {\max }\limits_{\textcolor{black}{\tilde{p} ,\tilde{q}} } \{  \tau_{p_{0 },q_{0 }} \left( t\right)  + \\ ~~~~~~~~~~~~~~~~~~~~\frac{\zeta_{\textcolor{black}{\tilde{p} ,\tilde{q}}} \left( t\right)  }{f_{\rm{c}}} +\frac{\psi_{p_{0 },q_{0 }} \left( t\right)  }{2\pi f_{\rm{c}}} \} \} -\mathop {\min }\limits_{p,q} \{ \tau_{p,q}( t)\}.   
    \end{array}
\end{equation}
In the following, we minimize the objective function in ${\textbf{\rm{OP}}}_{\boldsymbol{2}}$. Based on \eqref{max power phase shift} and the causality requirement, ${\psi}_{p,q}(t) \geq 0$, we have
\vspace{-4mm}
\begin{equation}\label{casuality}
\begin{array}{l}
     \zeta_{\textcolor{black}{\tilde{p} ,\tilde{q}}} \left( t\right)  \geq  
      -f_{\rm{c}}\varpi_{\textcolor{black}{\tilde{p} ,\tilde{q}}}(t)-\frac{\psi_{p_{0},q_{0}}(t)}{2\pi},
\end{array}
\end{equation}
from \eqref{casuality} and since $\zeta_{\textcolor{black}{\tilde{p} ,\tilde{q}}} \left( t\right)\in \mathbb{Z}$, the minimum value of $\zeta_{\textcolor{black}{\tilde{p} ,\tilde{q}}} \left( t\right)$ can be obtained as $\zeta^{\text{min}}_{\textcolor{black}{\tilde{p} ,\tilde{q}}} \left( t\right)=
       \left\lceil -f_{\rm{c}}\varpi_{\textcolor{black}{\tilde{p} ,\tilde{q}}}(t)-\frac{\psi_{p_{0},q_{0}}(t)}{2\pi}  \right\rceil$
which is a decreasing function with respect to $\psi_{p_{0},q_{0}}(t)$. Equation \eqref{Delay Spread Upper Bound after OP2} includes additional increasing function, i.e., $\frac{\psi_{p_{0},q_{0}}(t)}{2\pi f_{\rm{c}}}$. 
 By substituting $\zeta^{\text{min}}_{\textcolor{black}{\tilde{p} ,\tilde{q}}} \left( t\right)$ into \eqref{Delay Spread Upper Bound after OP2}, it is obvious that the variation of $\psi_{p_{0},q_{0}}(t)\in[0,2\pi]$ results in a small variation, less than $\frac{1}{f_{\rm{c}}}$, in  $T_{\rm{De}}^{\rm{upp}}\left( t\right)$. 
 Hence, we relax $\zeta^{\text{min}}_{\textcolor{black}{\tilde{p} ,\tilde{q}}} \left( t\right)$ to $\zeta^{\rm{R}}_{\textcolor{black}{\tilde{p} ,\tilde{q}}} \left( t\right)=\left\lceil -f_{\rm{c}}\varpi_{\textcolor{black}{\tilde{p} ,\tilde{q}}}(t) \right\rceil$, which turns \eqref{Delay Spread Upper Bound after OP2} into an increasing function with respect to 
 $\psi_{p_{0},q_{0}}(t)$. Accordingly, the closed-form solution for the RIS phase shifts are obtained as \eqref{proposed solution} by considering $\psi_{p_{0},q_{0}}(t)=0$ and substituting $\zeta^{\rm{R}}_{\textcolor{black}{\tilde{p} ,\tilde{q}}} \left( t\right)$ into \eqref{max power phase shift}. This closed-form solution is Pareto optimal based on Th. 4.2.1 in \cite{MOO-book}. Accordingly, \eqref{proposed solution} jointly optimizes \eqref{received power} and \eqref{doppler}, as the first priority order, and minimizes \eqref{delay spread upper bound} as the second priority order. 
\textcolor{black}{Another potential solution of ${\textbf{\rm{OP}}}_{\boldsymbol{}}$ is to reverse the priority order between ${\textbf{\rm{OP}}}_{\boldsymbol{1}}$ and ${\textbf{\rm{OP}}}_{\boldsymbol{2}}$. This reversed priority approach leads to non-efficient solution that is presented later in Section~\ref{simulation result}.}}
\end{proof}
\vspace{-2mm}
\textbf{\textit{Corollary 1.}}
\textit{With this Pareto optimal solution, the Doppler spread is zero, the maximum value for the instantaneous cascade channel gain is achieved as $\frac{P_{\rm{R}}^{\rm{max}}\left( t\right)}{P_{\rm{T}}} =\left| \sum\limits_{p=1}^{P}\sum\limits_{q=1}^{Q} \Gamma_{p,q} \left( t\right)  \right|^{2}$, and the delay spread upper bound is
\vspace{-2mm}
\begin{equation}\label{min delay spread}
\begin{array}{l}
       T_{\rm{De}}^{\rm{upp,min}}\left( t\right)=\max \{ \tau_{p_{0},q_{0}}(t) ,\mathop {\max }\limits_{\textcolor{black}{\tilde{p} ,\tilde{q}} } \{ \tau_{p_{0},q_{0}}(t)+
     \frac{\zeta^{\rm{R}}_{\textcolor{black}{\tilde{p} ,\tilde{q}}} \left( t\right)  }{f_{\rm{c}}}\} \}\\~~~~~~~~~~~~~~-\mathop {\min }\limits_{p,q} \{ \tau_{p,q} \left( t\right)\}.
\end{array}
\end{equation}
\textcolor{black}{
As can be seen in Fig.~\ref{RIS mobility pattern}, all the circular paths of the RIS elements have the same center located exactly between the Tx and Rx. With this symmetrical feature, the proposed closed-form solution works for all time slots due to the periodicity of the mobility patterns of the RIS elements.}
}
\vspace{-4mm}
\section{Numerical Evaluations}\label{simulation result}
In this section, we evaluate the performance of our proposed RIS phase shift design in Section~\ref{sec: RIS phase shift design}.
We assume a circular path with the origin $(0,0,0)$ and the radius $R_0=3$ km parallel to $xy$-plane. The RIS dimensions are chosen in a way such that $a=20\times b$, i.e., the length is much larger than the width. This is because the RIS is mounted below the HAPS wing, as in Fig.~\ref{sys}.
The RIS element dimensions are chosen as $d_{x}=d_{y}=\frac{\lambda_{\rm{c}}}{5} $, where $f_{c}=2$ GHz, and hence, the total number of RIS elements can be obtained as $P\times Q=\lceil \frac{a}{d_{x}} \rceil \times \lceil \frac{b}{d_{y}} \rceil =\lceil \frac{5a}{\lambda_{\rm{c}} } \rceil \times \lceil \frac{a}{4\lambda_{\rm{c}} } \rceil$. 
HAPS altitude and velocity of 20~km and $v=110$~km/h are used in our simulations, respectively. These parameters are inline with the specifications of one of the well-known aerodynamic HAPS, HAWK30, \cite{Halim-HAPS-Types, HAPS-Mobile}. The terrestrial Tx and Rx coordinates in the scale of km are $(x_{\text{T}},y_{\text{T}},z_{\text{T}})=(-5,0,20)$ and $(x_{\rm{R}},y_{\rm{R}},z_{\rm{R}})=(5,0,20)$, respectively. The planar antenna gain of RIS element $(p,q)$ to $S$ can be considered as $g^{S }_{{}p,q}\left( \theta^{S }_{p,q} \left( t\right)  ,  \varphi^{S }_{p,q} \left( t\right)  \right)=\frac{4\pi}{\lambda^{2}_{c}}d_{x}d_{y}\cos\theta^{S }_{p,q}(t)$ for $\theta^{S  }_{p,q}(t)\in \left[0,\frac{\pi}{2}\right]$ and zero otherwise, \cite{Emil WCL}.
The gains of the transmit and receive antennas are $g_{S }^{p,q}(t)=1$.
As mentioned in \textit{Lemma 2}, an alternative approach to our proposed method is optimization with reversed priority, i.e., reversing the order of ${\textbf{\rm{OP}}}_{\boldsymbol{1}}$ and ${\textbf{\rm{OP}}}_{\boldsymbol{2}}$ in the optimization process. Hence, in the following, we compare our proposed method with this alternative approach.
   \begin{figure}[t!]
    \subfigure[]{\includegraphics[width=0.236\textwidth]{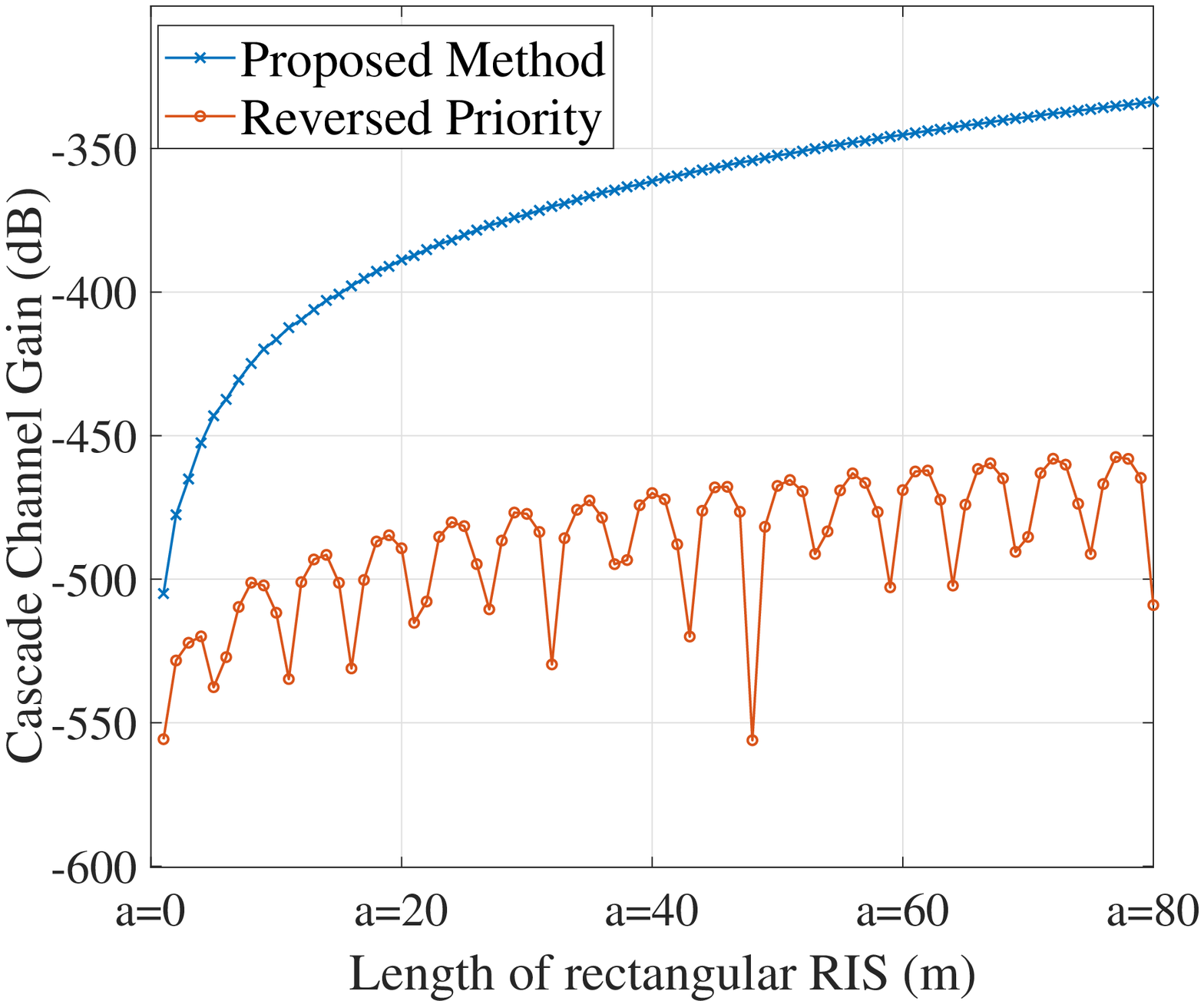}\label{Compare phase shift}} 
        \hspace{-4mm}
    \subfigure[]    
    {\includegraphics[width=0.25\textwidth]
    {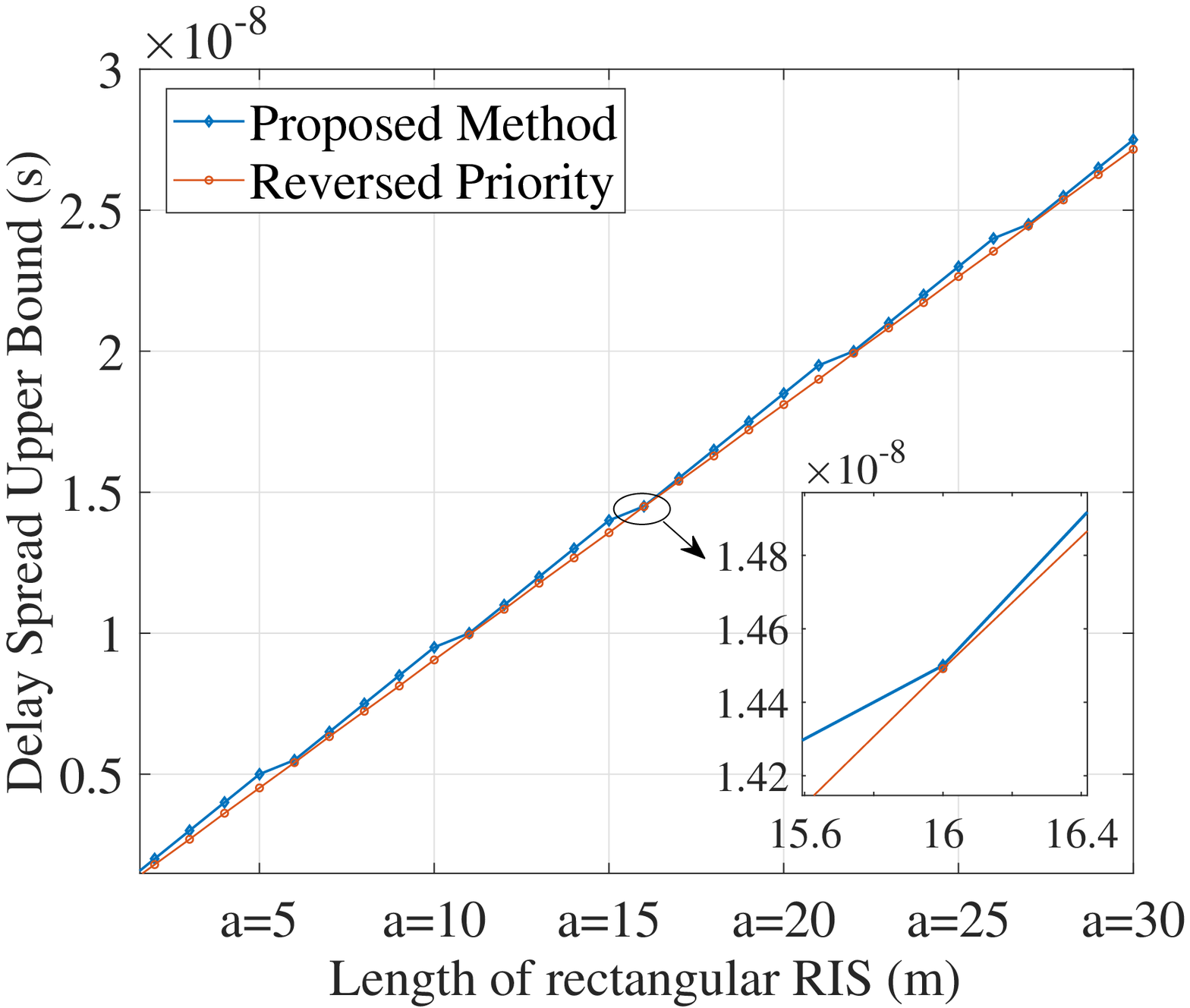}\label{Compare phase shift, delay spread}} 
    \vspace{-4mm}
    \caption{(a) Cascade channel gain versus RIS dimensions at $t=t_0$. (b) Delay spread upper bound versus RIS dimensions at $t=t_0$. }
    \label{fig:foobar}
\end{figure}                                
\begin{figure}[t!]
    \vspace{-3mm}
    \subfigure[]{\includegraphics[width=0.255\textwidth]{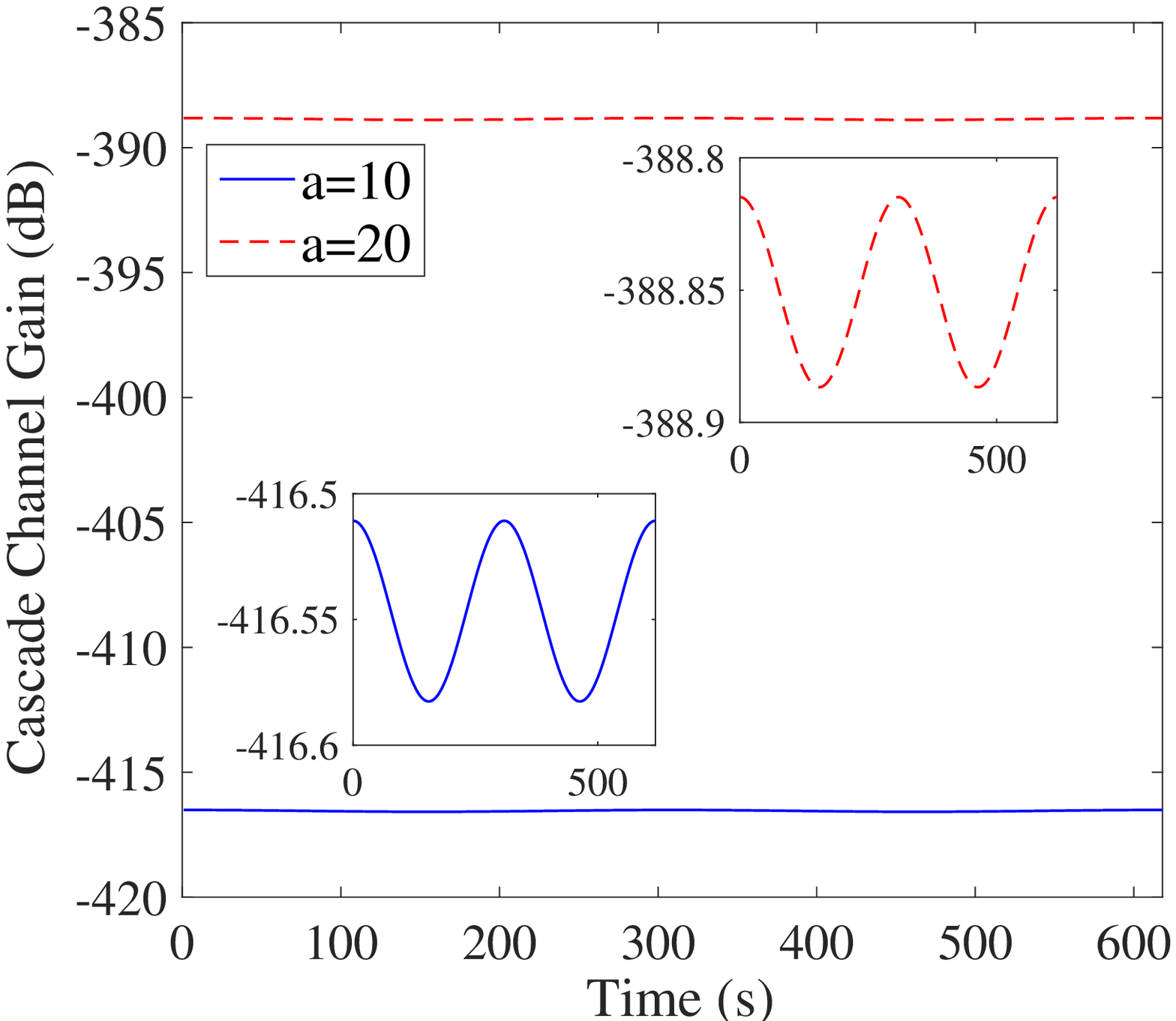}\label{Channel gain vs time}} 
    \hspace{-6mm}
    \subfigure[]{\includegraphics[width=0.25\textwidth]{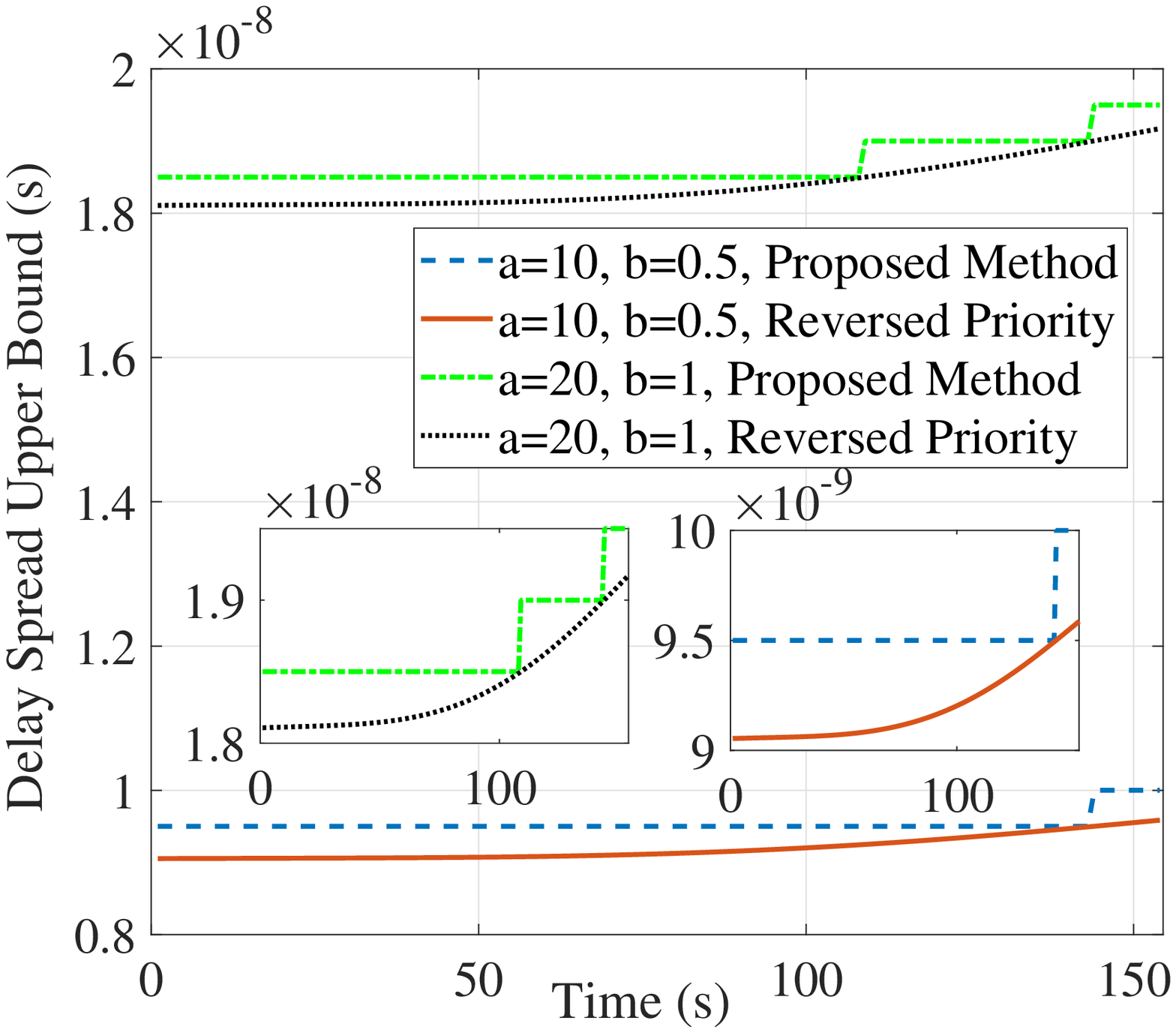}\label{DSupp vs time}} 
    \caption{(a) Cascade channel gain versus time for different RIS dimensions. (b) Delay spread upper bound versus time for different RIS dimensions.}
    \label{fig:foobar}
\end{figure}    
In Fig.~\ref{Compare phase shift} and Fig.~\ref{Compare phase shift, delay spread}, the cascade channel gain and $T_{\rm{De}}^{\rm{upp}}(t)$ of the proposed method and reversed approach are compared at a snapshot $t_0=10$~s. 
 Using the result of \textit{Corollary~1}, in Fig.~\ref{Compare phase shift}, we plot the cascade channel gain versus RIS dimensions. As can be seen, in Fig.~\ref{Compare phase shift}, the reversed approach leads to a poor performance compared to our proposed method.
 Exploiting the proposed method makes the cascade channel gain controllable and it can be constructively increased by increasing the RIS dimensions.
In contrast, the cascade channel gain based on reversed approach is uncontrollable as the only controllable parameter, i.e., the RIS phase shifts are fixed. This is due to the fact, mentioning in \textit{Lemma 2}, that $\psi_{p,q}(t)$ is considered zero $\forall p,q,t$ to optimize $T_{\rm{De}}^{\rm{upp}}(t)$ with the first priority order. By substituting $t=t_0$ s and zero phase shifts in \eqref{received power}, the cascade channel gain can be formulated as $\frac{P_{\rm{R}}(t_0)}{P_{\text{T}}}= \left|\sum\limits_{p=1}^{P}\sum\limits_{q=1}^{Q} \Gamma_{p,q}(t_0) \exp(-j2\pi f_{\rm{c}}\tau_{p,q}( t_0))\right|^{2}$.  
 The term $\exp(-j2\pi f_{\rm{c}}\tau_{p,q}( t_0))$ can negatively affect the cascade channel gain and makes it uncontrollable. Adopting the results of \textit{Corollary 1}, in Fig.~\ref{Compare phase shift, delay spread}, we plot $T_{\rm{De}}^{\rm{upp}}(t_0)$ versus RIS dimensions to compare the proposed method and the reversed one. It can be seen that the delay spread gap between our proposed method and the reversed priority is negligible. Furthermore, by extrapolating Fig. \ref{Compare phase shift, delay spread}, we can see that for $a=80$ m, $T_{\rm{De}}^{\rm{upp}}(t_0)$ is around $8\times 10^{-8}$~s. This is due to the almost linear behavior of $T_{\rm{De}}^{\rm{upp}}(t_0)$ as a function of $a$.
 Therefore, \eqref{proposed solution} can keep the delay spread upper bound controllable even for a large number of RIS elements. The claims for Fig.~\ref{Compare phase shift} and Fig.~\ref{Compare phase shift, delay spread} are feasible 
 for any $t=t_0$ based on the results presented in  Figs.~\ref{Channel gain vs time} and \ref{DSupp vs time}.
 In Figs.~\ref{Channel gain vs time} and \ref{DSupp vs time}, we analyze the cascade channel gain and $T_{\rm{De}}^{\rm{upp}}(t)$ versus time for different RIS dimensions, respectively. Fig. \ref{Channel gain vs time} shows that by increasing the value of $a$ from 10~m to 20~m, the cascade channel gain can be increased by 27.7~dB. \textcolor{black}{As can be seen, the fluctuation is less than 0.1~dB and can be ignored as it is negligible compared to the average value of the cascade channel gain. There is no significant benefit to consider the time-varying transmit signal to compensate this negligible fluctuation.}
In Fig.~\ref{DSupp vs time}, we plot $T_{\rm{De}}^{\rm{upp}}(t)$ versus time to compare our proposed method with the reversed approach. As can be seen, the gap is less than $5\times10^{-10}$~s and it is negligible. In addition, it is clear that \eqref{proposed solution} can make $T_{\rm{De}}^{\rm{upp}}(t)$ controllable for different time slots. 
 \vspace{-4mm}
 \section{Conclusion}
 \textcolor{black}{In this letter, we proposed a new network architecture exploiting an aerodynamic HAPS-RIS to provide connection between the unconnected ground stations. We proposed a multi-objective optimization problem for designing the RIS phase shifts based on the predictable mobility of aerodynamic HAPS-RIS. We found a closed-form solution for the RIS phase shifts, adopting Pareto optimality, based on an RIS reference element. We maximized the channel gain, forced the Doppler spread to zero, and minimized the delay spread upper bound. 
By exploiting this closed-form Pareto optimal solution, we do not need to constantly track the channel variations and constantly update the RIS phase shifts by solving optimization problems. Finally, we showed the performance efficacy of our proposed closed-form solution through numerical simulation. }
 

\vspace{-7mm}
 \begin{thebibliography}{9}
  	\bibitem{3GPP release} 
 “Study on new radio to support non-terrestrial networks,” 3GPP,
 Sophia Antipolis, France, 3GPP Rep. TR 38.811, 2018. [Online].
 Available: https://www.3gpp.org/ftp//Specs/archive.



 \bibitem{NTN-RIS-alouini} 
	J. Ye, J. Qiao, A. Kammoun, and M.S. Alouini, ``Non-terrestrial communications assisted by reconfigurable intelligent surfaces,"
	\textit{Proc. of the IEEE}, vol. 110, no. 9, pp. 1423-1465, 2022.
 
 	\bibitem{NTN-RIS-jamalipour} 
	P. Ramezani, B. Lyu, and A. Jamalipour, ``Toward RIS-enhanced integrated terrestrial/non-terrestrial connectivity in 6G," \textit{IEEE Network}, pp. 1-9, 2022.
	
		\bibitem{Halim-HAPS-UAV-SAT} 
	S. Alfattani, W. Jaafar, Y. Hmamouche, H. Yanikomeroglu, A. Yongacoglu, N. D. D\`{a}o, and P. Zhu, ``Aerial platforms with reconfigurable smart surfaces for 5G and beyond," \textit{IEEE Commun. Mag.}, vol. 59, no. 1, pp. 96-102, 2021.

	
    \bibitem{Halim-link-budget} 
	S. Alfattani, W. Jaafar, Y. Hmamouche, H. Yanikomeroglu, and A. Yongaçoglu, ``Link budget analysis for reconfigurable smart surfaces in aerial platforms," \textit{IEEE Open Journal of the Commun. Soc.}, vol. 2, pp. 1980-1995, 2021.
   \textcolor{black}{\bibitem{RIS-Relay} 
   E. Björnson, Ö. Özdogan, and E. G. Larsson, "Intelligent reflecting surface versus decode-and-forward: How large surfaces are needed to beat relaying?." 
   \textit{IEEE Wireless Commun. Lett.},
    vol. 9, no. 2, pp. 244-248, 2019.}
	
		\bibitem{Halim-HAPS-Types} 
	G. Kurt, M. Khoshkholgh, S. Alfattani, A. Ibrahim, T. Darwish, M. Alam, H. Yanikomeroglu, and A. Yongaçoglu, ``A vision and framework for the high altitude platform station (HAPS) networks of the future,"
	\textit{IEEE Commun. Surv. Tut.}, vol. 23, no. 2, pp. 729-779, 2021.
	
	
	\bibitem{HAPS-RIS-ICC} 
 S. Alfattani, A. Yadav, H. Yanikomeroglu, A. Yongacoglu, “Beyond-cell communications via HAPS-RIS,” in \textit{Proc. 2022 IEEE Globecom Workshops
(GC Wkshps)}, 2022, pp. 1383–1388.


 
 	\bibitem{HAPS-RIS-efficient} 
S. Alfattani, A. Yadav, H. Yanikomeroglu, A. Yongacoglu, “Resource-efficient HAPS-RIS enabled beyond-cell communications,” \textit{IEEE Wireless Commun. Lett.}, 2023.
\textcolor{black}{\bibitem{High Mobility RIS}
 C. Pan et al., “An overview of signal processing techniques for RIS/IRS-aided wireless systems,” \textit{IEEE J. Sel. Topics Signal Process.}, vol. 16, no. 5, pp. 883–917, Aug. 2022.}

\bibitem{CE-RIS-TV} 
 S. Sun and H. Yan, “Channel estimation for reconfigurable intelligent surface-assisted wireless communications considering Doppler effect,” \textit{IEEE Wireless Commun. Lett.}, vol. 10, no. 4, pp. 790-794, Apr. 2021.

 \bibitem{RIS-Roadside-Rzhang} 
Z. Huang, B. Zheng, and R. Zhang, “Roadside IRS-aided vehicular
communication: efficient channel estimation and low-complexity
beamforming design,” \textit{arXiv preprint arXiv:2207.03157}, Jul. 2022.

 \bibitem{Emil WCL} 
	B. Matthiesen, E. Björnson, E. De Carvalho, and P. Popovski, ``Intelligent reflecting surface operation under predictable receiver mobility: A continuous time propagation model,"
	\textit{IEEE Wireless Commun. Lett}, vol. 10, no. 2, pp. 216-220, 2020.
 
\bibitem{CE-MobileRIS-TV} 
 W. Wu, H. Wang, W. Wang, and R. Song, “Doppler mitigation method
aided by reconfigurable intelligent surfaces for high-speed channels,”
\textit{IEEE Wireless Commun. Lett.}, vol. 11, no. 3, pp. 627–631, 2022.


\bibitem{CE-MobileRIS-TWC} 

 Z. Huang, B. Zheng, and R. Zhang, “Transforming fading channel from
fast to slow: Intelligent refracting surface aided high-mobility communication,” \textit{IEEE Trans. Wireless Commun.}, vol. 21, no. 7, pp. 4989–5003,
Jul. 2022.

\bibitem{RIS-SAT-JSAC} 
B. Zheng, S. Lin, and R. Zhang, “Intelligent reflecting surface-aided
LEO satellite communication: Cooperative passive beamforming and distributed channel estimation,” \textit {IEEE JSAC}, vol. 40, no. 10, pp. 3057–3070,
Oct. 2022.

 \bibitem{Emil-RIS-Dimensions}
 Ö. Özdogan, E. Björnson, and E. G. Larsson, “Intelligent reflecting surfaces: Physics, propagation, and pathloss modeling,” \textit{IEEE Wireless Commun. Lett.}, vol. 9, no. 2, pp. 581–585, May 2020.

 \bibitem{emil mag} 
	E. Bj\"{o}rnson, H. Wymeersch, B. Matthiesen, P. Popovski, L. Sanguinetti, and E. Carvalho ``Reconfigurable intelligent surfaces: A signal processing perspective with wireless applications,"
	\textit{IEEE Signal Process. Mag.}, vol. 39, no. 2, pp. 135-158, 2022.

 \bibitem{Friis} 
H. T. Friis, “A note on a simple transmission formula,” \textit{Proc. of the IRE}, vol. 34, no. 5, pp. 254–256, 1946.

	
	 \bibitem{Emil-MOO} 
E. Björnson, E. A. Jorswieck, M. Debbah, and B. Ottersten, ``Multi-objective signal processing optimization: The way to balance conflicting metrics in 5G systems,"
\textit{IEEE Signal Process. Mag.}, vol. 31, no. 6, pp. 14–23, 2014.
	
	\bibitem{MOO-book} 
	K. Miettinen, \textit{Nonlinear Multiobjective Optimization}. Boston, MA, USA: Springer, 1999.
	
	\bibitem{HAPS-Mobile} 
	HAPSMobile. Accessed: Nov. 24, 2022. [Online]. Available: https://www.hapsmobile.com/


	
\end{thebibliography}
\end{document}